\documentclass{mem}
\usepackage{natbib}\usepackage{txfonts}\usepackage{balance}
\usepackage{graphicx}
\usepackage[a4paper]{hyperref}
\idline{0}{1}
\begin{document}
\def\teff{$T\rm_{eff }$}
\def\kms{$\mathrm {km s}^{-1}$}

\title{
High-amplitude $\delta$~Scuti stars in the Galactic bulge from the OGLE-II and MACHO data
}

   \subtitle{}

\author{
A.\,Pigulski,
Z.\,Ko{\l}aczkowski,
T.\,Ramza
\and A.\,Narwid
}

\offprints{A.\,Pigulski}
 
\institute{
Instytut Astronomiczny Uniwersytetu Wroc{\l}awskiego,
Kopernika 11, 51-622 Wroc{\l}aw, Poland,
\email{pigulski@astro.uni.wroc.pl}
}

\authorrunning{Pigulski et al.}

\titlerunning{HADS in the Galactic bulge}

\abstract{Searching for main-sequence pulsators, we analyzed photometry of $\sim$200,000 
variable star candidates from the OGLE-II Galactic fields, 
finding 193 high-amplitude $\delta$~Scuti stars.  This doubles the number of
known stars of this type.  The MACHO data, available for half 
of stars, were also analyzed.  In our sample of the HADS stars, we found
50 multiperiodic objects, including 39 that have period ratios in the range of 
0.76--0.80, an indication of the radial fundamental and first-overtone pulsation.  
We discuss the resulting Petersen diagram for these stars in view of the period 
ratios predicted by models.  Except for stars showing pulsations in
the radial fundamental mode and first overtone, we find the evidence for higher radial
overtones and non-radial modes in the analyzed sample of multiperiodic HADS stars.

}
\maketitle{}

\section{Introduction}
During the last several years, large photometric time-series data 
have been made available.  Using massive photometries
from the OGLE, MACHO, ASAS and NSVS projects, we started a systematic
study of the Galactic main-sequence pulsators.  This paper presents the
results of the search for high-amplitude $\delta$ Scuti (hereafter HADS) stars
in the catalogue of $\sim$200,000 variable star candidates published by \citet{wozn02}.
This catalogue contains $I$-filter photometry carried out between 1997 and 2000 within the 
OGLE-II project \citep{udal97}. The observations covered about 11 square degrees in 49 Galactic fields.
The OGLE-II data were supplemented by the MACHO observations \citep{alak01}, if available.

HADS stars form a subgroup of $\delta$ Scu\-ti stars with
large amplitudes (exceeding $\sim$0.3~mag in $V$ band), non-sinusoidal light curves
and small rotational velocities (see, e.g., \citealt{mcna97}).  There is, however, no clear separation
between them and $\delta$~Scuti stars with small amplitudes.  HADS stars pulsate mainly in radial modes, but there
is a growing evidence that non-radial modes with small amplitudes can also be unstable
in these stars.  Population II HADS stars are called SX Phoenicis stars and are
discovered mainly in the globular clusters.  At present, we know about 150 HADS stars in the
Galactic field \citep{rodr00} and about 200 SX Phe stars in globular clusters and nearby galaxies
(\citealt{rolo00}, \citealt{clem01}).

The MACHO microlensing survey data in Galactic fields were already used to search for the HADS stars by \citet{alco00}, who
found 90 stars of this type, including 18 multiperiodic objects. 

\section{Analysis}

The first two steps in our analysis were done automatically.  All stars from the
catalogue of variable star candidates of \citet{wozn02} were analyzed by means of
the Fourier periodogram.  Consecutive detection and subtraction of up to five periodic terms
was performed.  The Fourier spectra were calculated up to 40~d$^{-1}$.  Next, the star 
that appeared to be variable was classified by means of the detected period(s), amplitude(s) and the results of
Fourier decomposition.  This information was used to select stars that were subsequently 
analyzed interactively.  At this step, it was decided whether the star was included in the
list of the HADS stars or not.  We adopted the following definition of a HADS star: it is
a star with period shorter than 0.25~d for which at least one harmonic of the main mode was
detected and which is not an RR Lyrae or W UMa star.  The latter two types were distinguished 
from the HADS stars by means of Fourier coefficients and/or visual inspection of the light
curves.   In total, 193 HADS stars were found.

It is interesting to mention that this procedure allowed us to discover also other main-sequence 
variables or at least candidates.  Many of them are multiperiodic.  We have found about 230 $\beta$~Cephei and low-amplitude 
$\delta$~Scuti stars (Narwid et al., these proceedings), and about 600 candidates for SPB and $\gamma$~Doradus stars.

As mentioned in the Introduction, for stars classified as HADS the photometry from MACHO 
database was retrieved, if present, and analyzed
in a similar way.  It turned out that about a half of HADS stars detected in the OGLE data have MACHO photometry,
but only 15 of them were found by \citet{alco00}. 
\begin{figure}[t!]
\resizebox{\hsize}{!}{\includegraphics[clip=true]{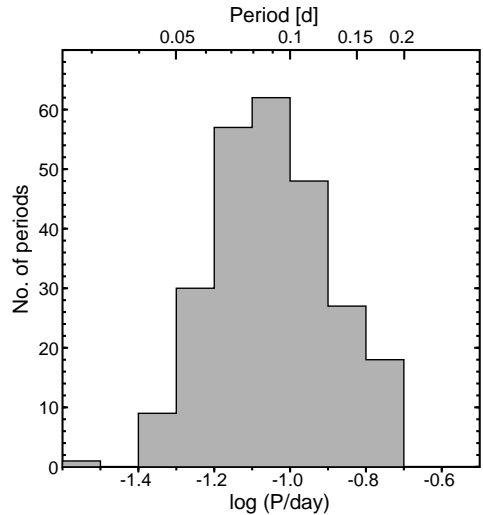}}
\caption{
\footnotesize
The histogram of periods detected in 193 HADS stars found in the OGLE-II
Galactic fields.
}
\label{histo}
\end{figure}
\begin{figure}[t!]
\resizebox{\hsize}{!}{\includegraphics[clip=true]{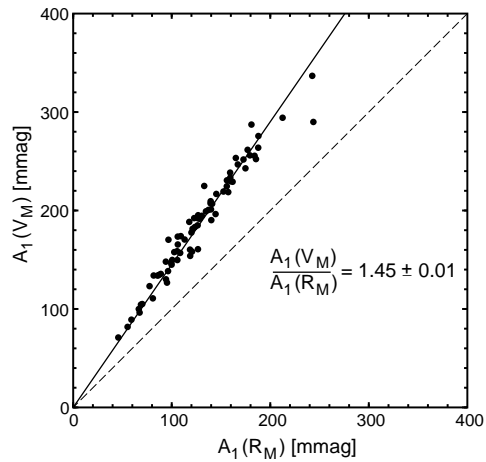}}
\caption{
\footnotesize
The semi-amplitudes of the main components of the highest-amplitude mode in the MACHO blue filter, $A_1(V_{\rm M})$, plotted
against semi-amplitudes of the component in the MACHO red filter, $A_1(R_{\rm M})$.  The continuous line is the best-fit of the
amplitude ratio equal to 1.45 $\pm$ 0.01.  The dashed line denotes $A_1(V_{\rm M})$ = $A_1(R_{\rm M})$. 
}
\label{ratio}
\end{figure}

\section{Results}

The histogram of periods detected in all 193 HADS stars is shown in Fig.~\ref{histo}.
We see that it peaks for periods equal to 0.08--0.1~d and that no single period longer than 0.2~was found. On the
other hand, the shortest period we detected equals to about 0.03~d.  For about 90 stars common to OGLE-II and MACHO data
we were able to find the ratios of amplitudes of the main modes.  The mean ratios of amplitudes were equal 
to $A_1(V_{\rm M})/A_1(I)$ = 1.80 $\pm$ 0.03, $A_1(R_{\rm M})/A_1(I)$ = 1.25 $\pm$ 0.02, and $A_1(V_{\rm M})/A_1(R_{\rm M})$ 
= 1.45 $\pm$ 0.01, where $A_1$ stands for the amplitude of the main component of the highest-amplitude
mode in the Fourier decomposition.  The MACHO blue and
red filters are denoted as $V_{\rm M}$ and $R_{\rm M}$, respectively.  The $A_1(V_{\rm M})$ vs.~$A_1(R_{\rm M})$ 
dependence is shown in Fig.~\ref{ratio}.  The small r.m.s.~errors of the ratios can be easily explained, since 
the modes detected in the HADS stars are mostly radial, and therefore their amplitudes should change with 
wavelength in a similar way. 
\begin{figure*}[t!]%
\resizebox{\hsize}{!}{\includegraphics[clip=true]{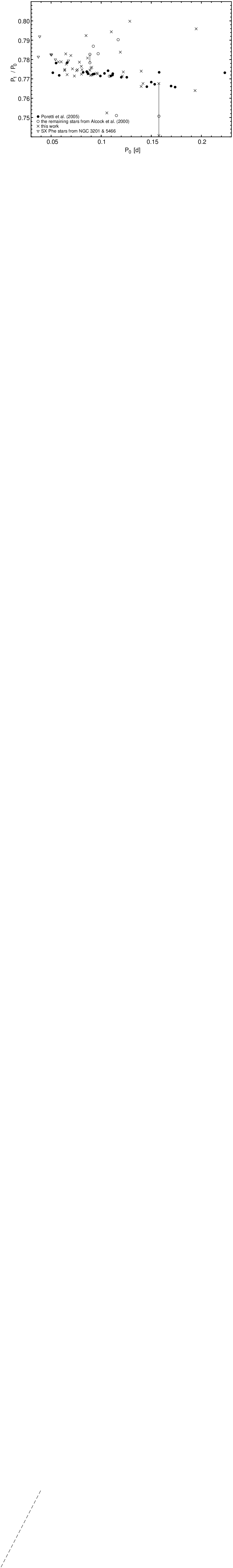}}
\caption{\footnotesize
Petersen diagram for double and multiple-mode HADS stars with period ratio in the range between 0.74 and 0.81.  
Filled circles: 25 stars listed by \citet{pore05}, 
open circles: the remaining seven stars discovered by \citet{alco00}, but not included by \citet{pore05}, crosses: 
39 stars from this study, open triangles: six double-mode SX Phe stars from NGC\,3201 and NGC\,5466.
Two vertical lines join points for the same star, but two possible values of $P_1$.
}
\label{peter}
\end{figure*} 

In our sample of 193 HADS stars, 52 stars show multiperiodic behaviour. Of them, 40 stars are double-mode pulsators,
while the remaining 12 have three or even more modes excited.  Double-mode HADS stars are very useful, as if the
period ratio falls in the range of 0.76--0.79, the modes can be interpreted as radial: fundamental and 
first overtone \citep{pech96}.

Double-mode HADS stars were recently studied by means of the Petersen diagram by \citep{pore05}, who provided an 
updated list of 25 known stars of this type.   These stars are shown with filled circles 
in Fig.~\ref{peter} which is the Petersen diagram
\footnote{In principle, the Petersen diagram is used to plot the ratio of the period of the first overtone
and fundamental radial modes, i.e. $P_1/P_0$, versus $P_0$. However, we use this diagram in a broader context, plotting
the ratio of a period of a given mode and the period of the main mode, presumably the fundamental radial one.
This does not mean that in each case we deal with two radial modes.  In fact, for some stars, we use the Petersen diagram
to claim the opposite (see text).}
showing all period ratios (with respect to the period of the main mode, $P_0$)
in the range between 0.74 and 0.81. 
Seven double-mode HADS stars in the Poretti et al.'s (2005) list were found in the MACHO data. However, there are seven more stars, 
not included by these autors, that have period ratios not very close to 0.77, but still in the range covered in Fig.~\ref{peter}.
These stars are shown with open circles in Fig.~\ref{peter}.  In fact, there is one star, MACHO 120.21785.976, with the reported 
period ratio of 0.71199 \citep{alco00}.  If, however, the alias frequency of the $f_2$ mode is taken as the true one, 
the period ratio goes very close to the canonical value and amounts to 0.77141.  We plot the latter value in Fig.~\ref{peter}.  

We add 39 stars to this sample (crosses in Fig.~\ref{peter}), which increases the number of known double-mode HADS stars
more than twofold.
The new stars populate mostly the short-period region of the diagram, where no star of this type was found by \citet{alco00}, 
apparently due to some selection effects in their analysis.

In order to explain the Petersen diagram for the HADS stars, \citet{pore05} used appropriate stellar models for different
masses and metallicities to calculate theoretical values of $P_1/P_0$.  A similar work has been done earlier by \citet{pech96}.
These calculations show that both mass and metallicity influences $P_1/P_0$.  The effect of mass is largest for periods longer 
than $\sim$0.12~d where lower mass leads to the lower value of $P_1/P_0$.  On the other hand, the metallicity is most important
in the short-period range.  The smaller metallicity the larger $P_1/P_0$.  The metallicity effect
is, in general, confirmed by the observations of double-mode SX Phe stars in globular clusters.  This can be
judged from Fig.~\ref{peter} where, for comparison, we show (open triangles) period ratios for six double-mode SX Phe stars
which are the members of two metal-poor Galactic globular clusters: NGC\,3201 \citep{mazu03} and NGC 5466 \citep{jeon04}.

Consequently, we may conclude that some double-mode HADS stars we found in the OGLE-II data could be SX Phe stars,
especially those that have periods shorter than 0.1~d and $P_1/P_0 > $ 0.775.  It is also obvious that
some of the ratios shown in Fig.~\ref{peter} fall outside the range of $P_1/P_0$ allowed by models.  There are at least
two plausible explanations of this fact.  For ratios larger than 0.78 we may deal with a pair of consecutive radial overtones 
(see \citealt{temp02}), radial and non-radial mode or even two non-radial modes.  
We see that for at least two stars shown in Fig.~\ref{peter}, this
must be true, as all their periods cannot be explained in terms of only radial modes.  

In addition to the modes shown in Fig.~\ref{peter}, we found in multiperiodic HADS stars modes that are good candidates 
for higher radial overtones and non-radial modes.
There is also an evidence for period changes in some HADS stars we found.  This conclusion comes from the fact that 
for these stars a peak at frequency very close to the main mode appears in the spectrum of residuals. 
 
\begin{acknowledgements}
This work has been supported by the KBN grant No.~1\,P03D\,016\,27.
\end{acknowledgements}

\bibliographystyle{aa}

\end{document}